\pdfoutput=1

\newcommand*{\ATLASLATEXPATH}{}
\documentclass[cernpreprint=true, texlive=2016, UKenglish,LANGEDIT=true]{\ATLASLATEXPATH atlasdoc}

\makeatletter
\DeclareOldFontCommand{\rm}{\normalfont\rmfamily}{\mathrm}
\DeclareOldFontCommand{\sf}{\normalfont\sffamily}{\mathsf}
\DeclareOldFontCommand{\tt}{\normalfont\ttfamily}{\mathtt}
\DeclareOldFontCommand{\bf}{\normalfont\bfseries}{\mathbf}
\DeclareOldFontCommand{\it}{\normalfont\itshape}{\mathit}
\DeclareOldFontCommand{\sl}{\normalfont\slshape}{\@nomath\sl}
\DeclareOldFontCommand{\sc}{\normalfont\scshape}{\@nomath\sc}
\makeatother

\usepackage{\ATLASLATEXPATH atlaspackage}
\usepackage{\ATLASLATEXPATH atlasbiblatex}
 
\usepackage{\ATLASLATEXPATH atlasphysics}
 
\graphicspath{{logos/}{figures/}}
 
\usepackage{ANA-STDM-2018-01-PAPER-defs}

\AtlasTitle{Measurement of differential cross sections for single diffractive dissociation
in $\sqrt{s} = \SI{8}{\TeV}$ $pp$ collisions using the ATLAS ALFA spectrometer}
 
\AtlasVersion{2.4 (18 October 2019)}
 
\AtlasAbstract{
A dedicated sample of Large Hadron Collider proton--proton collision
data at centre-of-mass energy $\sqrt{s}=\SI{8}{\TeV}$
is used to
study inclusive single diffractive dissociation, $pp \rightarrow Xp$. The intact final-state
proton is reconstructed in the
ATLAS ALFA forward spectrometer, while charged particles from
the dissociated system $X$ are measured in the central detector components.
The fiducial range of the measurement is $-4.0 < \log_{10} \xi < -1.6$ and
$0.016 < |t| < 0.43 \ {\rm GeV^2}$, where $\xi$ is the
proton fractional energy loss and $t$ is the squared
four-momentum transfer. The total cross section integrated across the fiducial range is
$1.59 \pm 0.13 \ {\rm mb}$.
Cross sections are also measured differentially
as functions of $\xi$,  $t$, and $\Delta \eta$, a variable that characterises the
rapidity gap separating the proton and the system $X$. The
data are consistent with an exponential $t$ dependence, ${\rm d} \sigma / {\rm d} t \propto \text{e}^{Bt}$
with slope parameter $B = 7.65 \pm 0.34 \ {\rm GeV^{-2}}$. Interpreted
in the framework of triple Regge phenomenology, the $\xi$ dependence leads to a pomeron intercept of
$\alpha(0) = 1.07 \pm 0.09$.}

\AtlasRefCode{ATLAS-STDM-2018-01}
 
\PreprintIdNumber{CERN-EP-2019-190}

\AtlasJournal{JHEP}
\AtlasJournalRef{JHEP 02 (2020) 042}
\AtlasDOI{10.1007/JHEP02(2020)042}

\AtlasCoverSupportingNote{Supporting Note}{https://cds.cern.ch/record/2300132}

\AtlasCoverEgroupEditors{atlas-stdm-2018-01-contact-editors@cern.ch}
 
\AtlasCoverEgroupEdBoard{atlas-stdm-2018-01-editorial-board@cern.ch}
 
\hypersetup{pdftitle={ATLAS document},pdfauthor={The ATLAS Collaboration}}
 
\begin{document}
 
\maketitle

\section{Introduction}

\begin{figure*}[h]
\centering
\subfloat[]{\includegraphics[width=0.32\textwidth]{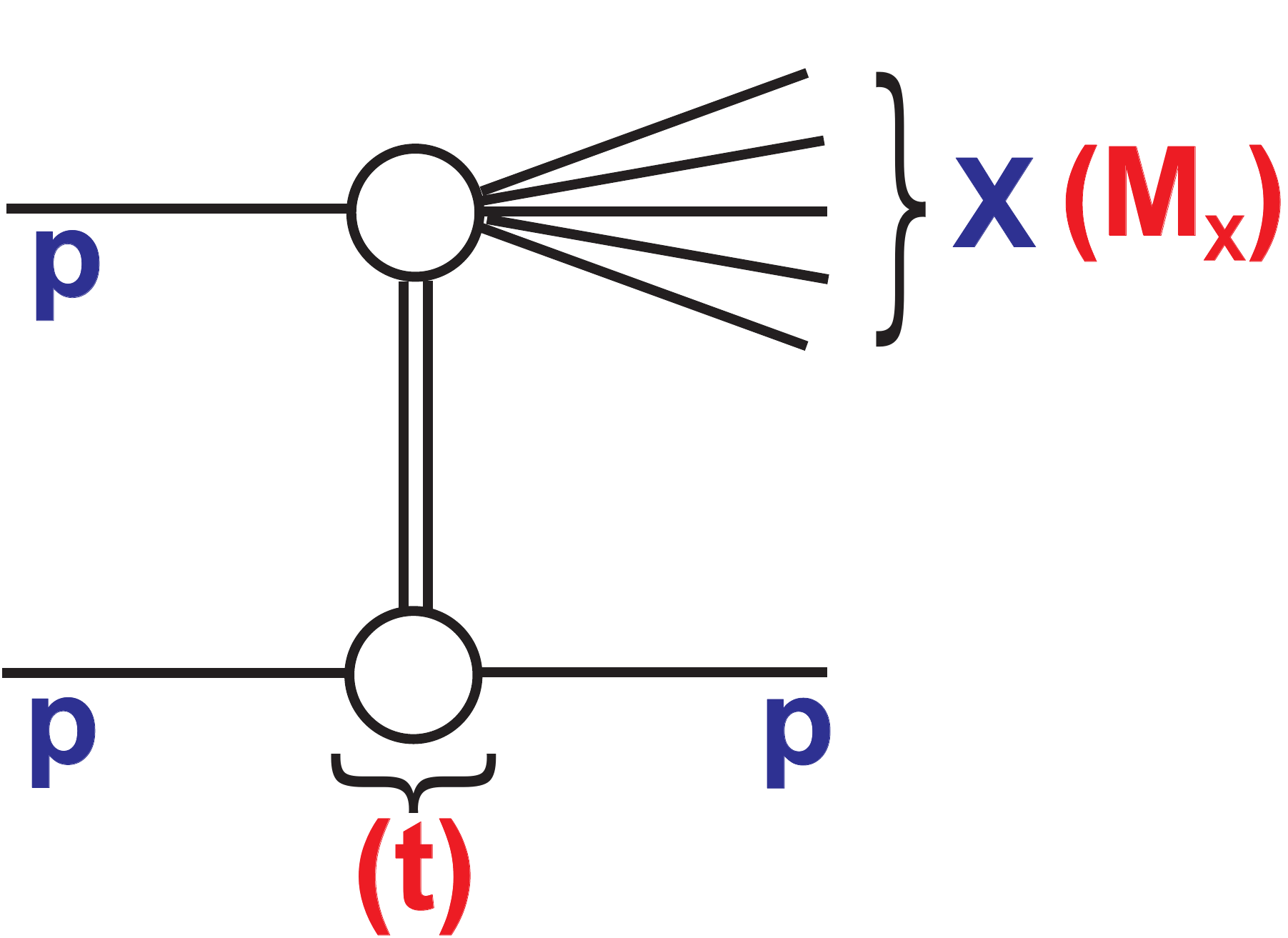}}
\hspace*{0.05\textwidth}
\subfloat[]{\includegraphics[width=0.3\textwidth]{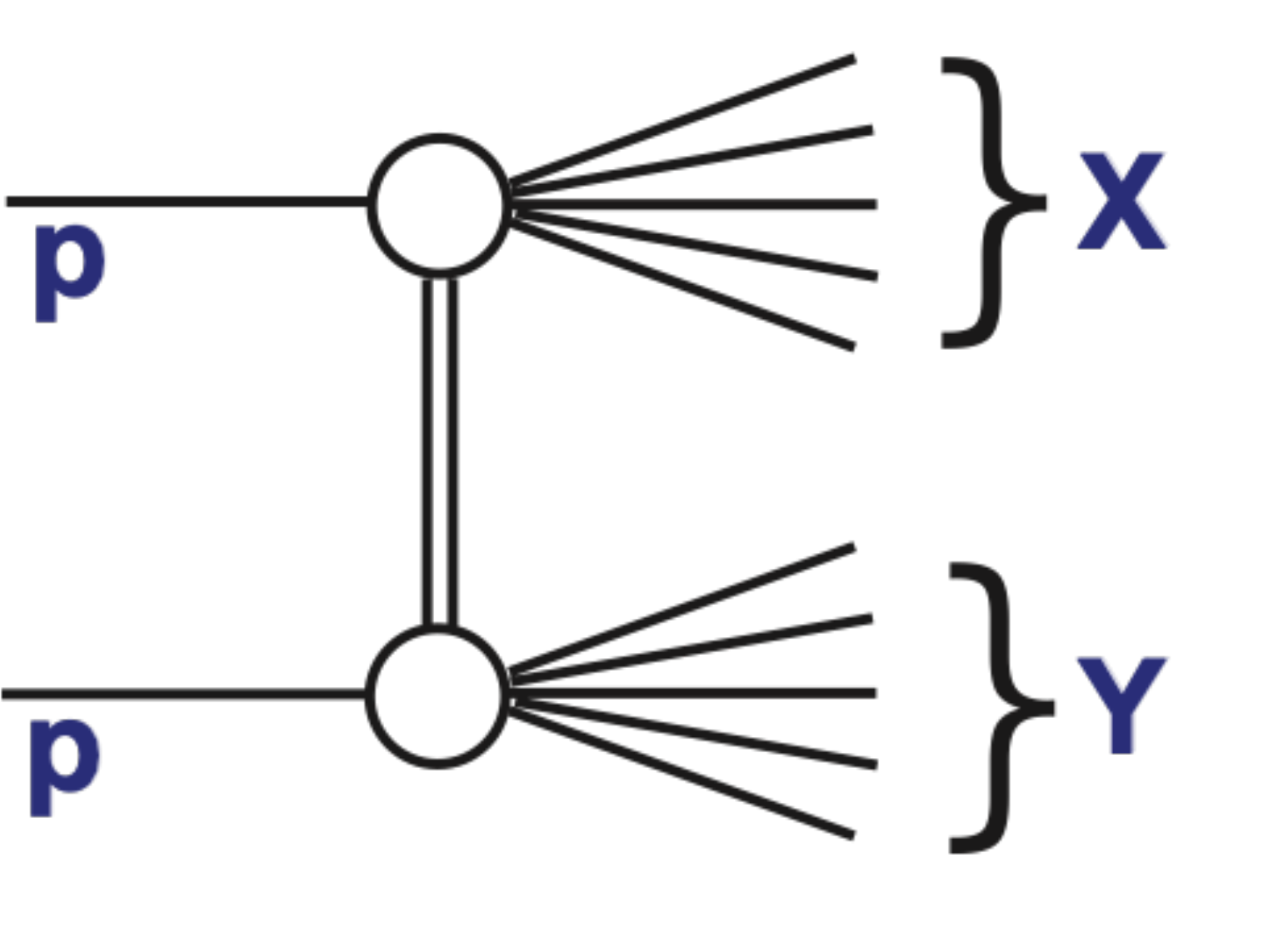}}
\hspace*{0.05\textwidth}
\subfloat[]{\includegraphics[width=0.28\textwidth]{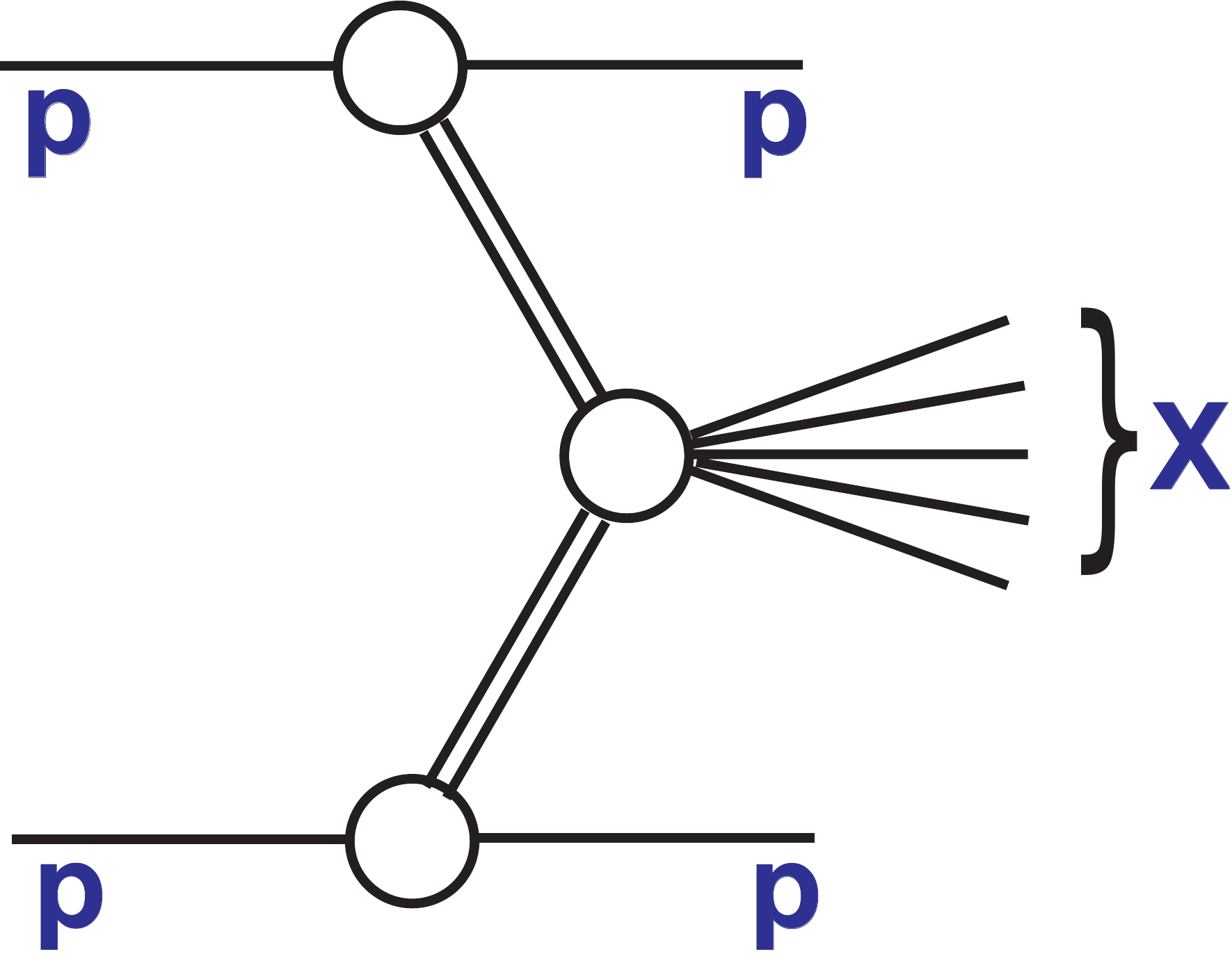}}
\caption[]{Schematic illustrations of the (a) single diffractive dissociation (SD),
(b) double diffractive dissociation (DD) and
(c) central diffraction (CD) processes. The kinematic variables used
to describe the SD process (the squared four-momentum transfer, $t$, and the mass, $M_X$, of the dissociated system $X$)
are indicated in parentheses in (a).}
\label{Feynmans}
\end{figure*}

In the single diffractive (SD) dissociation process in proton--proton ($pp$) collisions, $pp \to Xp$ (Figure~\ref{Feynmans}(a)),
the absolute value of the squared four-momentum transfer $t$ is usually much smaller than
$1 \ {\rm GeV^2}$, such that the intact final-state proton is scattered through a very small angle
of typically $10$--$100 \ {\rm \mu rad}$.
The other proton dissociates to produce a multi-particle hadronic
system $X$, whose mass $\mx$ can reach many
hundreds of ${\rm GeV}$ at Large Hadron Collider (LHC) energies,
whilst remaining in a regime where the fractional energy loss of the intact
proton
$\xi = \mx^2 / s$
is small.
 
Measurements of the SD cross section have been made
at a wide range of energies
\cite{GOULIANOS1983169,ALBERI19811,0038-5670-31-2-R02,Kaidalov:1979jz},
most recently at the SPS \cite{Bernard:1986yh,Ansorge:1986xq},
the Tevatron \cite{Amos:1992jw,Abe:1993wu} and HERA \cite{Newman:2013ada}.
The process is usually interpreted
phenomenologically in terms
of the exchange of a strongly interacting net colour-singlet, sometimes referred to as a
pomeron \cite{Feinberg1956}.
The range of applicability of a universal pomeron across
total, elastic, and diffractive processes
has a long history of investigation.
Despite the wealth of previous data, predictions for the SD contribution
at the CERN Large Hadron Collider
(LHC) vary widely. Our current lack of constraints limits the precision
of direct measurements of the total inelastic $pp$
cross section \cite{Aad:2011eu}.
Diffraction is also an important ingredient
in understanding the low Bjorken-$x$
region of proton structure \cite{Newman:2013ada} and cosmic-ray air
showers \cite{Ostapchenko:2016wtv}, and it
may even be related to the string theory of gravity \cite{Brower:2006ea}.
 
Cross sections related to diffractive dissociation have been measured using early LHC
data \cite{Aad:2012pw, Kha:2015gka,Abelev:2012sea}
by exploiting the `large rapidity gap' signature that is kinematically
expected. Whilst they clearly establish the presence of a large
diffractive contribution, these measurements are not able to distinguish
fully between the SD process, its
double dissociation (DD, $pp \rightarrow XY$, Figure~\ref{Feynmans}(b))
analogue in which both protons dissociate, and the tail of non-diffractive
(ND) contributions
in which large rapidity gaps occur due to random fluctuations in the hadronisation process.
The large rapidity gap measurements also
do not offer direct access to the underlying dynamics in $\xi$ and $t$.
 
This paper reports a measurement of the SD process in
which the intact final-state proton is reconstructed, suppressing DD and ND contributions to
negligible levels and allowing a study of the
cross section differentially in $t$.
The cross section is also measured differentially in $\xi$ as
obtained from the reconstructed charged-particle tracks in the ATLAS
central detector
and in $\Delta \eta$,
a variable characterising the size of the central
pseudorapidity
region in which no charged
particles are produced.
 
 
\section{Experimental conditions}
\newcommand{\AtlasCoordFootnote}{
ATLAS uses a right-handed coordinate system with its origin at the nominal interaction point
in the centre of the detector and the $z$-axis along the beam-pipe.
The $x$-axis points from the interaction point to the centre of the LHC ring,
and the $y$-axis points upwards.
Cylindrical coordinates $(r,\phi)$ are used in the transverse plane,
$\phi$ being the azimuthal angle around the $z$-axis.
The pseudorapidity is defined in terms of the polar angle $\theta$ as $\eta = -\ln \tan(\theta/2)$.
Angular distance is measured in units of $\Delta R \equiv \sqrt{(\Delta\eta)^{2} + (\Delta\phi)^{2}}$.}

ATLAS is a multipurpose apparatus covering almost the entire solid angle around
its LHC collision
point~\cite{PERF-2007-01}.\footnote{\AtlasCoordFootnote}
This measurement makes use of the sensitivity of the inner tracking detector (ID) and
the minimum-bias trigger scintillators (MBTS) to the components of the dissociating system $X$.
 
The ID consists of a high-granularity silicon pixel detector from which the
vertex location is reconstructed, surrounded by a silicon microstrip particle tracker.
These silicon detectors are complemented by a transition radiation
tracker,
and are enclosed within a \SI{2}{\tesla} axial magnetic field, enabling precise
charged-particle tracking in the range $|\eta| < 2.5$.
The MBTS detectors are mounted on the front faces of the calorimeter endcaps on
both sides of the interaction point and cover the pseudorapidity range $2.1 < |\eta| < 3.8$.
They consist of two  concentric discs of scintillating tiles,
each segmented in azimuth into eight counters.

The ALFA forward proton spectrometer \cite{AbdelKhalek:2016tiv}
consists of vertically oriented `Roman pot station' insertions to the beam-pipe
at \SI{237}{\metre} and
\SI{241}{\metre} from the interaction point on both sides of ATLAS, housing movable scintillating fibre detectors.
At each station, detectors approach the
beam from above and below (i.e.\ in the $y$ direction).
There are thus four `armlets', each of which consists of a pair
of detectors either above or below the beam on one side of the interaction
point, from which proton tracks can be reconstructed.
The main detectors consist of 20
layers of 64 fibres each, arranged in 10 overlapping pairs in two perpendicular ($u, v$) orientations at
$45^\circ$ to the ($x,y$) coordinates,
read out by an array of
multi-anode photomultiplier tubes.
These main detectors are supplemented by scintillating tiles, which provide trigger signals.
For the run studied, the
innermost parts of the sensitive detectors
were placed at 9.5$\sigma$ of the beam envelope, corresponding to $7.5 \ {\rm mm}$ from the beam centre.
 
The data sample used in this analysis was taken during a
dedicated data-taking period in July 2012, which has also been used to measure the elastic and
total cross sections at $\sqrt{s} = \SI{8}{TeV}$ \cite{Aaboud:2016ijx}.
The luminosity was kept very low by LHC standards, such that the
mean number of inelastic interactions per bunch crossing
(`pile-up')
is never more than 0.08, allowing rapidity gaps to be identified
and suppressing random coincidences between protons in ALFA and unrelated activity in the central
detector components. The data were taken in a high-$\beta^{*}$
quadrupole configuration, which
provides beams of almost collinear protons at zero crossing angle.
This allows the sensitive components of ALFA to be placed as close as possible to the beam,
enabling detection of protons at very small deflection angles.
 
The best estimate of the track position in a Roman pot is given
by the overlap region of the hit areas of all fibres,
which leads to a local precision of around $30 \ {\rm \mu m}$ in each coordinate.
Correlating hits between pots and reconstructing the proton kinematics
relies on an alignment procedure, which is carried out using
elastic-scattering and beam-halo data \cite{Aaboud:2016ijx}.
A special beam optics configuration~\cite{Cavalier:1382028} was in place for the data used here,
incorporating `parallel-to-point' focusing in the vertical plane, such that
the $y$ coordinate of the proton impact point in a Roman pot detector at fixed $z$
depends only on the scattering angle and the energy loss.
The optics configuration does not provide the parallel-to-point feature
in the horizontal plane, so the $x$ coordinate
in the Roman pot detectors depends on the primary vertex position, which is measured by the central detector.
The combination of the $x$ and $y$ coordinates of the
signals in the Roman pot stations at $z = 237 \ {\rm m}$ and $z = 241 \ {\rm m}$
and the primary vertex are therefore used together to reconstruct
the values of $\xi$ and $t$.


\section{Monte Carlo simulation}
\label{sec:MC}
Monte Carlo (MC) simulations are used
for the modelling of background contributions, unfolding
of instrumental effects, and comparisons of models with the hadron-level cross-section measurements. The
\pythia~\cite{Sjostrand2008852} generator
was used to produce the main SD, ND and DD samples and also that for the `central diffractive'
(CD, $pp \rightarrow pXp$, Figure~\ref{Feynmans}(c)) process. The SD, DD and CD models in \pythia{} are
based on the exchange of a pomeron with trajectory $\alpha(t) = \alpha(0) + \alpha^\prime t$, assuming
`triple Regge' \cite{1970Mueller} formalism (see Section~\ref{sec:result}).
The models~\cite{PhysRevD.49.2257} are tuned using previous ATLAS data, including
the total inelastic cross section \cite{Aad:2011eu} and
rapidity gap spectra \cite{Aad:2012pw}.
By default, the `A3' tune \cite{ATL-PHYS-PUB-2016-017} was used, which adopts the
`Donnachie--Landshoff' \cite{Donnachie:1992ny} choice for the pomeron flux factor
to describe the $\xi$ and $t$ dependences in the diffractive channels with
pomeron intercept $\alpha(0) = 1.07$.
An alternative SD sample was produced using the A2 tune \cite{ATL-PHYS-PUB-2012-003}
and the Schuler--Sj\"{o}strand model for the pomeron flux factor \cite{PhysRevD.49.2257},
which has $\alpha(0) = 1$ and therefore differs from Donnachie--Landshoff mainly in its $\xi$ dependence.
Both tunes use the
H1 2006 Fit B diffractive parton densities \cite{Aktas:2006hy}
as an input to model the hadronisation in the diffractive channels.
For the non-diffractive channel, the A3
tune uses the
NNPDF23LO \cite{Ball:2012cx}
proton parton densities.
Generated central particles were propagated
through the \geant{}~based simulation
of ATLAS~\cite{Agostinelli2003250,Aad:2010ah} to produce the simulated signals in the central detector components.
The generated protons in diffractive processes are transported from the
interaction point to the ALFA detectors
by representing each element of the LHC optical lattice
(quadrupole and dipole magnets) as a simple matrix under the thin-lens approximation,
giving the total transfer matrix once multiplied together.
 
The impact of uncertainties in the hadronisation properties of the dissociation
system $X$ is evaluated by comparison of \pythia{} with the cluster-based approach in the \herwig{}
Monte Carlo model \cite{Bahr:2008pv,Bellm:2015jjp}
(Version 7.1.3 is used).
\herwig{} makes predictions for
the diffractive cross section based on an updated model
of soft and diffractive processes \cite{Gieseke:2016fpz}, which adopts a triple Regge approach
and $\xi$ and $t$ parameterisation similar
to that in \pythia{} with the A3 tune, but produces final-state dissociation particles
according to a multi-peripheral model \cite{Amati:1962nv}.
 
\section{Data selection and reconstruction}
\label{sec:data}
Events are triggered by requiring activity in at least two MBTS counters
on the same side of the interaction point,
in coincidence with
a signal in a pair of `near' and `far' planes in
ALFA on the opposite side.
The efficiency of the
trigger is determined separately for each measurement interval
by reference to a randomly seeded trigger with the subsequent requirement of an ID
track with
transverse momentum
$\pt > 200 \ {\rm MeV}$,
corresponding to the minimum offline selection requirement in this analysis.
The trigger efficiency was cross-checked by replacing the reference trigger with one based on
the LUCID forward detector \cite{Aaboud:2016hhf}
and also with a sample triggered on completely random
bunch crossings.
After accounting for prescales, the integrated luminosity of the sample is $1.67 \pm 0.03 \ {\rm nb^{-1}}$,
as determined from van der Meer scans \cite{Aaboud:2016ijx,Aaboud:2016hhf}.
 
For the triggered sample, the
MBTS response is analysed segment-by-segment
with offline thresholds set to best separate signal from noise generated
in the photomultipliers and by the read-out electronics.
Thresholds are set individually for the 32 counters by fitting the noise distribution
around zero to a Gaussian distribution and placing the threshold at 4$\sigma$
from the Gaussian mean. The same approach is applied in the simulation.
The trigger efficiency rises relatively slowly with the number of active MBTS
segments according to this offline reconstruction.
Events are therefore required to have at least five MBTS counters
passing the offline requirements, at which point the trigger efficiency is approximately 50\%.
 
Events are required to have at least one good-quality charged-particle track reconstructed
in the ID as well as a reconstructed primary vertex.
The selection applied for the good-quality tracks follows
the criteria established in Ref. \cite{Aad:2010ac} and requires
$\vert \eta \vert<2.5$ and $p_{\rm T}>200 \ {\rm MeV}$ as well as the presence of hits in both the pixel
and strip detectors, an acceptable track fit
$\chi^2$ and
transverse and longitudinal track impact parameters relative to the nominal interaction point that are compatible
with a primary vertex.
The vertex reconstruction efficiency is very close to 100\% for events with four or
more reconstructed charged particles, falling off at lower
multiplicities \cite{Aad:2016xww}.
 
The ALFA alignment procedures \cite{Aad:2014dca,Aaboud:2016ijx}
lead to a precision at the level of $20$--$30 \ {\rm \mu m}$
for proton track segments in each Roman pot station.
Segments are reconstructed from hits
in at least six $u$ and six $v$ fibre layers.
To avoid areas of reduced performance close to the detector edges and
efficiency losses in the shadow of beam collimators, track segments are restricted in the
$y$ coordinate to a region
about $8 \ {\rm mm}$ to $20 \ {\rm mm}$
from the beam-line,
varying slightly between stations.
More than one track segment is reconstructed in an ALFA armlet in less than 1\% of cases;
the segment with the most overlapping fibres is then selected.
Proton tracks are reconstructed from the combination of segments above or below the beam
in adjacent near and far stations.
Further constraints are derived from the expected
correlation pattern between the average $x$-position of
the track segments in the near and far stations ($\bar{x}$)
and the local angle the track makes in the $(x,z)$
plane, $\theta_x$.
The region with low values of $\bar{x}$ and $\theta_x$ is populated dominantly by
SD processes at modest $\xi$, whereas
beam-related `halo' background contributions cover a wide range in $\bar{x}$ and backgrounds from non-SD
$pp$ collisions are
relatively
evenly distributed in both variables.
A bivariate Gaussian distribution is fitted to the
observed two-dimensional ($\bar{x}, \theta_x$) distributions for each armlet,
and tracks are accepted if they lie within a 3$\sigma$ contour of the resulting ellipse.
Only events with exactly one reconstructed proton track are considered in the analysis.
 
The intrinsic reconstruction efficiency of ALFA for minimum-ionising particles was
determined to be close to 100\% in test beams \cite{Aad:2014dca}. However, reconstruction
inefficiencies arise from failures of the track reconstruction algorithm,
mostly due to hadronic interactions.
The ALFA track reconstruction efficiency is obtained
separately for each armlet through a
`tag and probe' approach
using a sample of elastic-scattering events,
following the method employed
in the ATLAS elastic-scattering measurement \cite{Aaboud:2016ijx},
adapted for the ALFA track and event selection used in this analysis. The efficiency is
91\%--94\% depending on the armlet, and is accounted for by
appropriately weighting
reconstructed events.

 
\section{Kinematic variables and fiducial region}
\label{sec:fiducial}
 
The measurement is performed differentially
in $t$, which is determined from the scattered proton's transverse momentum as
reconstructed using ALFA. The resolution in $t$ is around $15 \%$.
 
The cross section is also measured differentially in the
`visible rapidity gap' variable, $\Delta\eta$. This variable represents the size of the region
in which no primary\footnote{A primary charged particle is defined
as a charged particle with a mean proper lifetime $\tau > 300 \ {\rm ps}$,
which is either directly produced in $pp$ interactions or from
decays of directly produced particles with
$\tau < 30 \ {\rm ps}$.}
charged particles are produced with $\pt > 200 \ {\rm MeV}$,
starting
at $|\eta| = 2.5$ on the same side of the
interaction point as the proton tag and extending towards the $X$ system.
This $\Delta\eta$ definition is similar
to that adopted in
Ref.~\cite{Aad:2012pw}, but is adapted to the current analysis, in which calorimeter
information is not used and charged particles are
reconstructed from tracks as described in Section~\ref{sec:data}.
The resolution in $\Delta \eta$ is relatively constant at
around 0.02.
 
The measurement is also performed as a function of $\xi$,
determined via $\xiS{}= \mx^2/s$
by using the
charged particles reconstructed in the ID to obtain the mass
of the diffractive system $X$.
The experimental sensitivity to $M_X$ is limited by the absence of
reconstructed neutral particles,
forward particles escaping the detector through the beam pipe, and low-$\pt$ particles
not reaching
the
detection threshold of the ID. The problem of the missing forward
particles is mitigated in the reconstruction by adopting
a similar approach to that in Ref.~\cite{Aad:2015xis}
which uses the approximation
$\xi \simeq \sum_{i}(E^{i} \pm p_{z}^{i}) / \sqrt{s}$. The corresponding reconstructed-level variable $\xi {\rm (ID)}$
is built from
the energies $E$ and longitudinal momenta $p_z$ of all measured ID tracks $i$, and
the sign $\pm$ is determined by the sign of the scattered proton's longitudinal momentum.
For this sum, the minimum requirement on track transverse momentum is relaxed from $200 \ {\rm MeV}$ to
$100 \ {\rm MeV}$. Missing neutral and remaining low-$\pt$ charged particles
are accounted for by applying a multiplicative linear function,
determined from the MC simulations, to the reconstructed $\log_{10} \xi$.
The ability of the simulations to provide this correction within the precision defined by the
associated systematics is supported by studies of charged particle distributions in diffraction
at the LHC~\cite{ATLAS-CONF-2010-048} and of diffractive charged particle spectra and total energy flows at previous colliders
such as HERA~\cite{Newman:2013ada}. Following this procedure,
the resolution is approximately constant in $\log_{10} \xi$ at around 0.3.
The variable $\xi$ can also be reconstructed using $\xi {\rm (ALFA)} = 1 - E_p^\prime / E_p$,
where $E_p^\prime$ and $E_p$ are the scattered proton's energy as measured by ALFA
and the beam energy, respectively.
Although the ALFA reconstruction has increasingly poor resolution as
$\xi$ becomes small,
it provides a powerful means of cross-checking the
ID-based measurement
with very different background
contributions, unfolding characteristics and other systematic effects.
 
The lower limit of the measurement
in $\xiS$ is determined by the inner detector and MBTS acceptance, while the
sensitive region in \tS{} and the upper limit in $\xi$ are determined
by the coverage of the ALFA stations.
The fiducial region is determined by consideration of the acceptance as evaluated
in the SD MC sample.
The acceptance in $\xi$ is approximately constant at around 30\%
over a wide range. The region $-4.0 < \log_{10} \xiS < -1.6$ is chosen,
for which the acceptance is at least half of the maximum value.
A fiducial range of $0.016 \ {\rm GeV^2} < |t| < 0.43 \ {\rm \GeV^2}$ is then taken, to ensure that the
acceptance is at least 10\% throughout the measured range.


\section{Backgrounds}
Background in the analysis arises from non-SD $pp$ collision processes leading to
correlated signals in ALFA
and the ID (`single-source'), as well as from coincidences
of a signal in ALFA with an uncorrelated signal in the ID (`overlay background').
 
The single-source contribution is dominated by the CD process, which naturally gives rise to
forward-going protons and activity in the ID. It is estimated using the
MC simulation,
reweighted through the comparison with data for the control sample described in
Section~\ref{sec:controlPlots}.
The probability that a \pythia{} CD event meets the selection criteria is 8.5\%.
The ND and DD single-source contributions are negligible.
 
In the overlay background, the signal in the central detector
almost always arises
from a ND, DD or SD $pp$ collision, whilst the ALFA signal may occur due to pile-up from
real forward-going protons in elastic-scattering or CD processes, showering
in DD or ND events, or from beam-induced
sources (mainly beam halo).
The overlay background is modelled using a data-driven technique in which the
normalisation is determined from the probability per bunch crossing
of reconstructing in ALFA a proton that passes the requirements applied in the main
analysis and is not correlated with central detector
activity. This probability is obtained
from a control data sample in which there are hits in all 32 MBTS segments
and reconstructed charged-particle tracks within 0.5 pseudorapidity units
of both edges of the ID acceptance. This large amount of
central detector activity
implies heavy suppression of
the phase space for correlated activity in ALFA.
There is a reconstructed proton in ALFA in
0.77\% of such events,
which sets the overlay background normalisation,
assuming that the ID and ALFA signals
are always uncorrelated. A $9 \%$ correction is made for
residual signal events in the sample, determined from MC simulations.
For the $t$ measurement, the shape of the overlay background contribution is taken directly
from the distribution in the control sample. For $\xi$ and $\Delta \eta$, the shape is
taken from the MC simulation of
ND, DD and SD events that pass the central detector requirements but do not
contain a proton in ALFA.
 
Background arising entirely from beam-induced processes or from `afterglow', in which
relics of previous events are recorded in a later bunch crossing, are studied using
monitoring samples from bunch crossings in which only one of the two proton beams is present and from sidebands in the $\left( \bar{x}, \theta_x \right)$ distribution. They contribute less than $0.1 \%$ of the total sample.


\section{Control distributions}
\label{sec:controlPlots}
\begin{figure}[p]
\centering
 
\subfloat[Nominal Sample]{
\includegraphics[width=.49\linewidth]{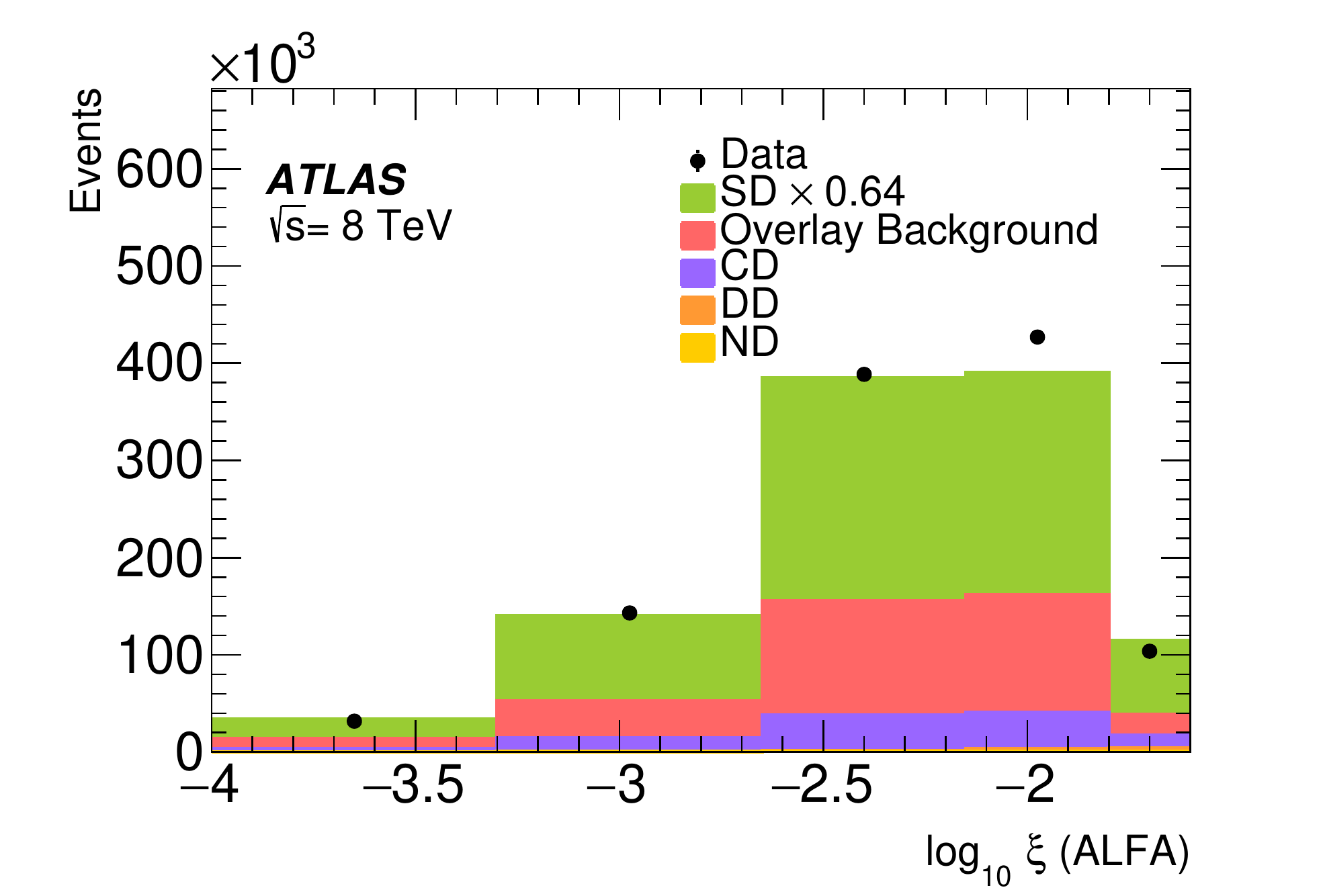}
}
\subfloat[Nominal Sample]{
\includegraphics[width=.49\linewidth]{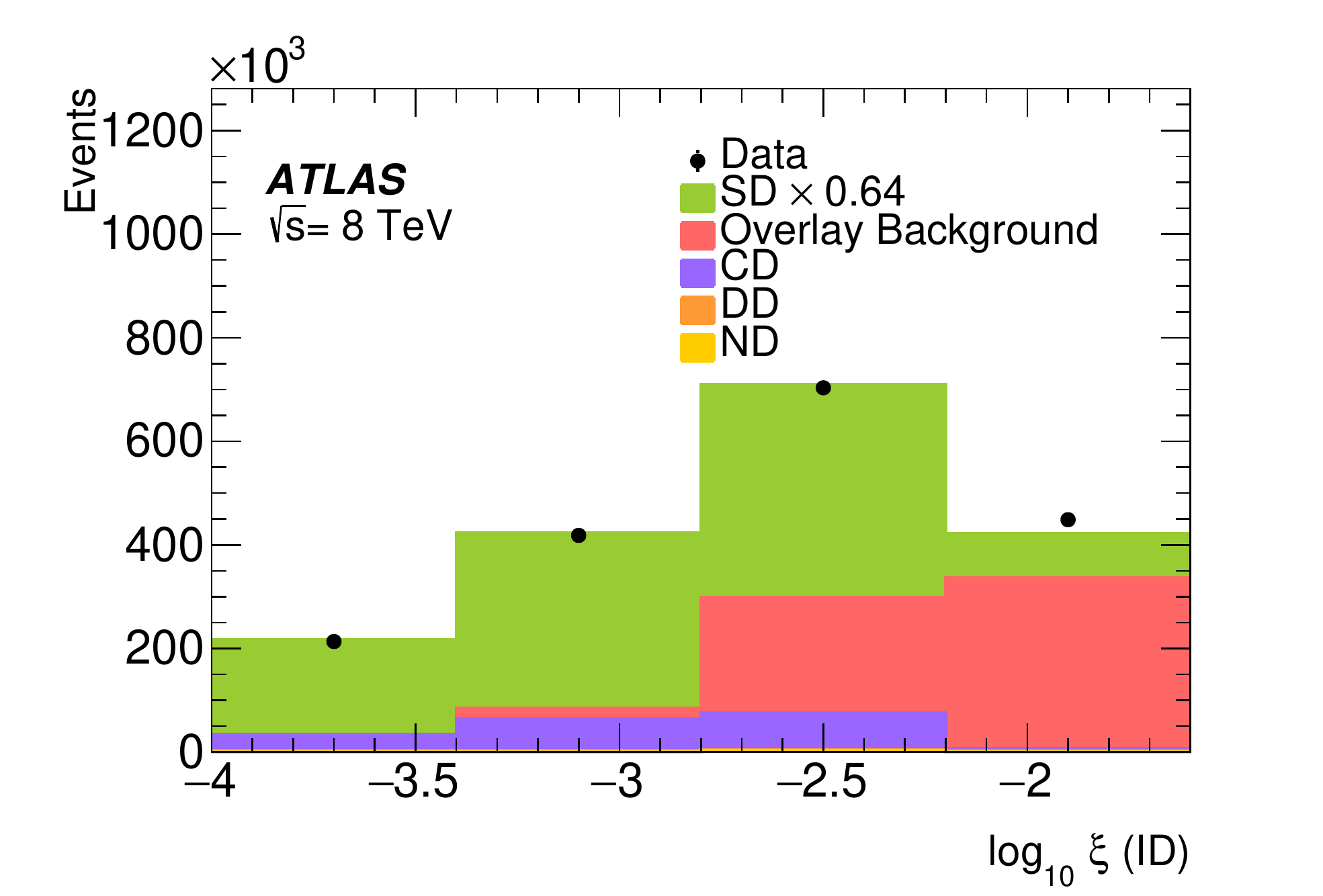}
}\hfil
\subfloat[Nominal Sample]{
\includegraphics[width=.49\linewidth]{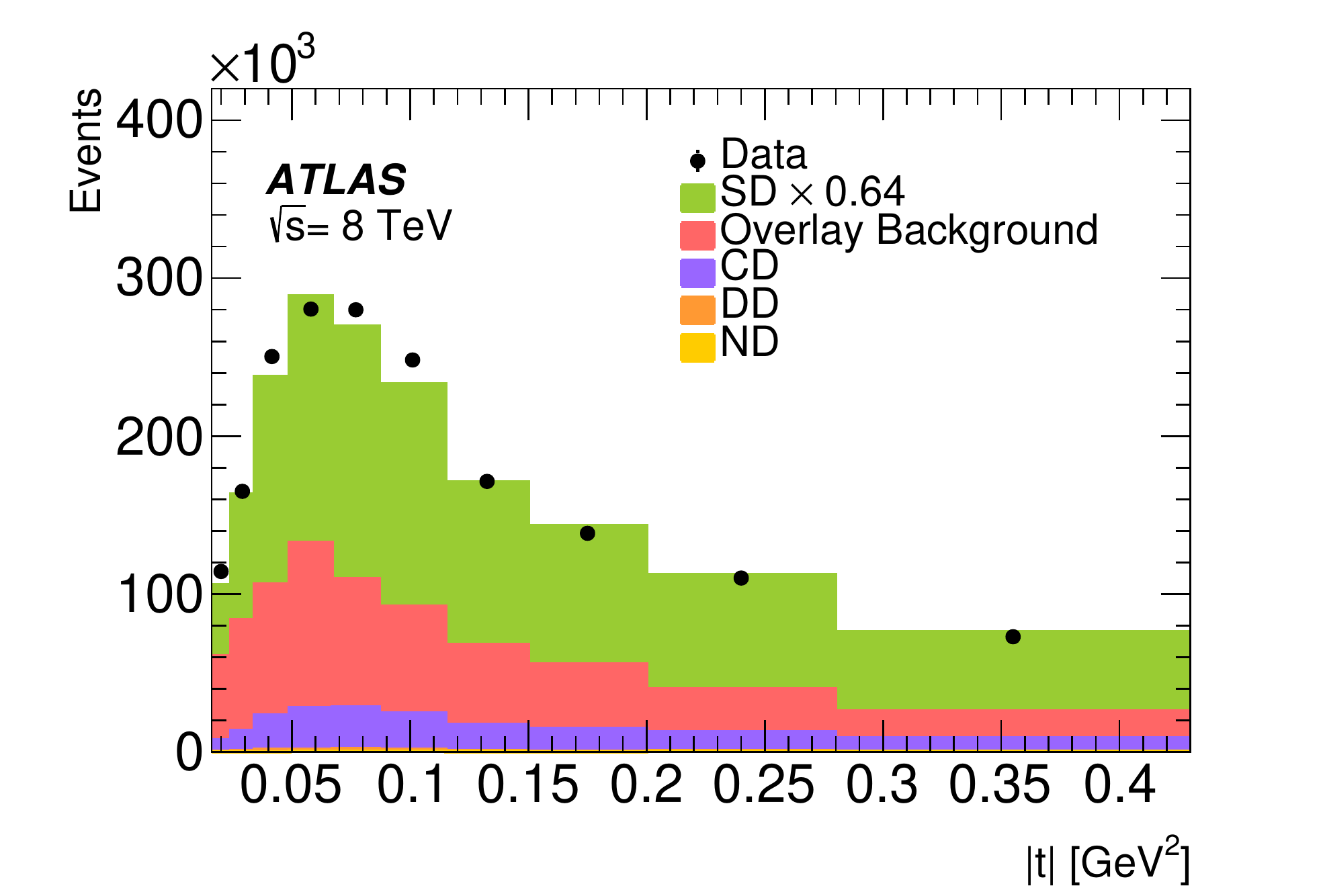}
}
\subfloat[Nominal Sample]{
\includegraphics[width=.49\linewidth]{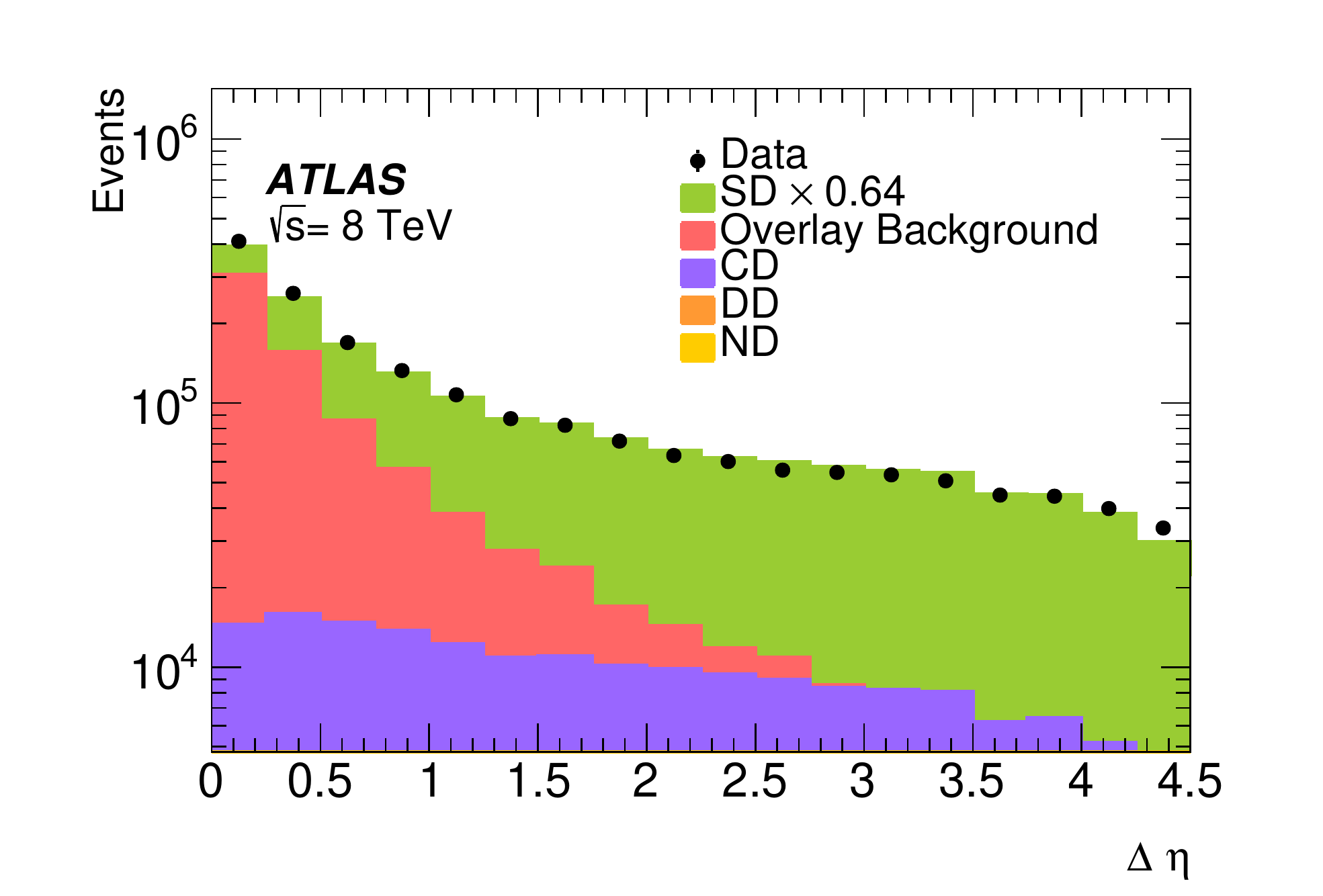}
}\hfil
\subfloat[Control Region 1]{
\includegraphics[width=.49\linewidth]{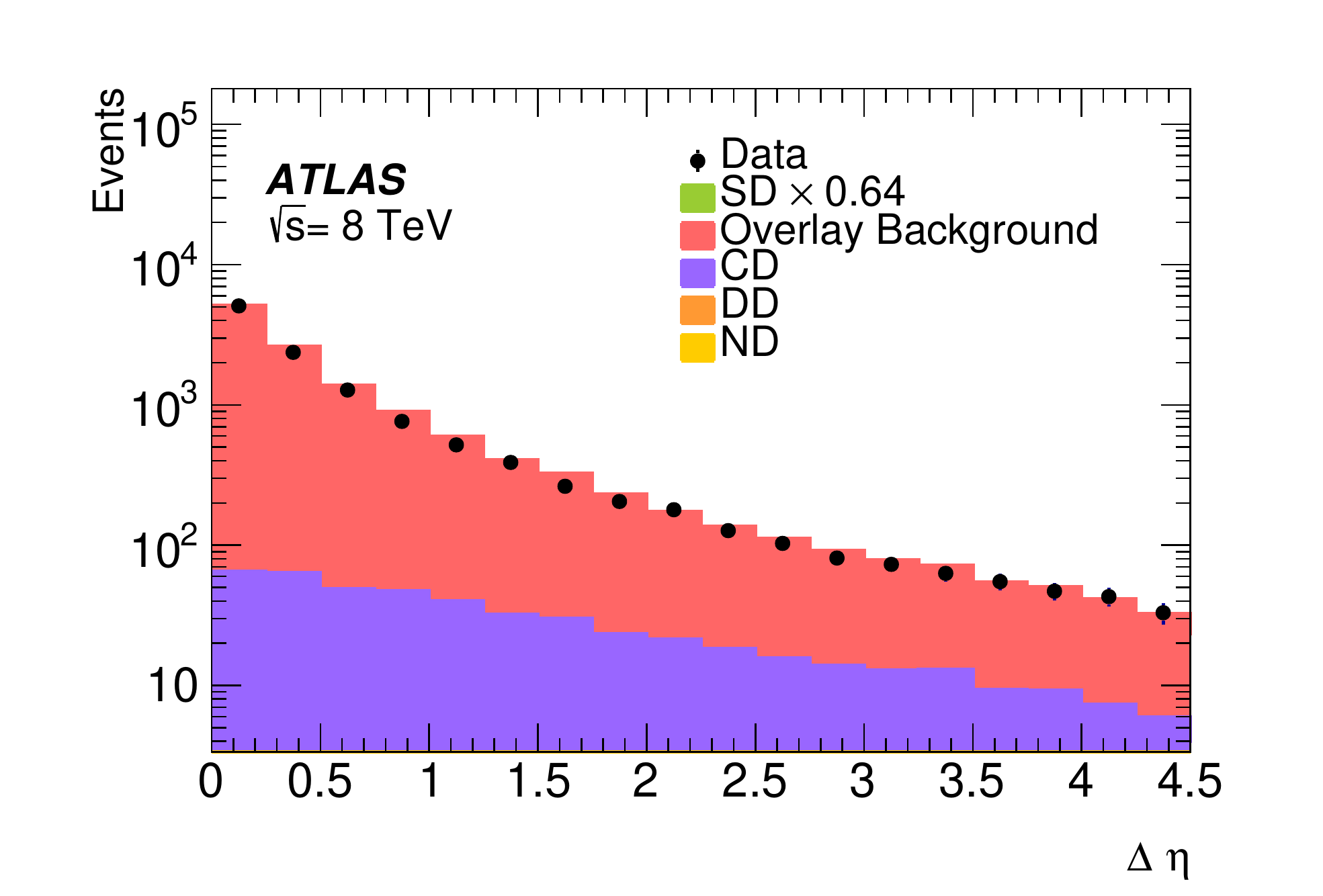}
}
\subfloat[Control Region 2]{
\includegraphics[width=.49\linewidth]{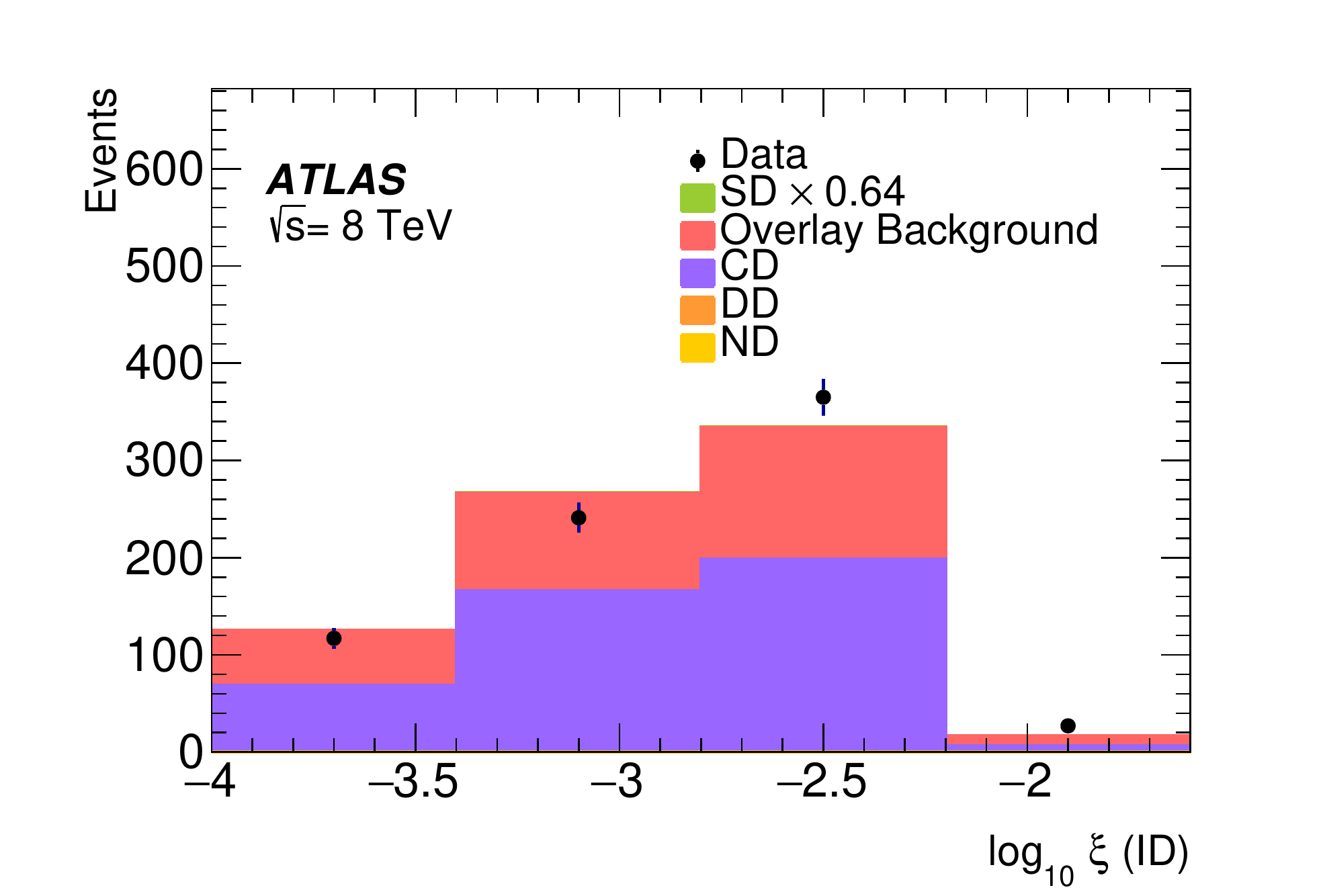}
}\hfil
\caption{Uncorrected (i.e. detector level) distributions of
(a) $\log_{10} \xi$ measured in ALFA, (b) $\log_{10} \xi$ mesaured in the ID, (c) $|t|$
and (d) $\Delta\eta$ for the basic selection of the measurement.
(e) Uncorrected $\Delta\eta$ distribution from Control Region 1, in which two proton
track segments are required rather than one.
(f) Uncorrected distribution in $\log_{10} \xi$ measured in the ID for Control Region 2,
in which exactly two proton track segments are required and the MBTS multiplicity is
required to be between 2 and 10.
In all distributions, data are compared with the sum of the
overlay background model and the \pythia~A3 tune prediction with
the SD contribution scaled by 0.64
to match the measurement in this paper.
In (f), the CD $\xi$ distribution at the MC generator level is reweighted as described in the text.
Significant contributions in (a) beyond the $\log_{10} \xi$ range of the measurement are not shown.
The uncertainties shown are statistical.
\label{fig:controlPlotsTuned}}
\end{figure}

Example control distributions, in which uncorrected data are compared with
predictions based on MC simulations and the data-driven background
model, are shown in Figure~\ref{fig:controlPlotsTuned}(a)--\ref{fig:controlPlotsTuned}(d). Here, the normalisations of the ND, DD and
CD MC models are taken from their \pythia{} default cross sections of
$51 \ {\rm mb}$, $8.3 \ {\rm mb}$ and $1.2 \ {\rm mb}$, respectively,
whilst the SD cross section in \pythia{} is adjusted from
$12.5 \ {\rm mb}$  to
$8.0 \ {\rm mb}$
to match the results of this analysis (see Section~\ref{sec:result}).
With these normalisations, all
variables are well described. The shape of the distribution in $|t|$ reflects the ALFA acceptance.
The SD contribution dominates in much of the phase space. The overlay
background contribution is largest at small
$\Delta \eta$, and at high values of $\xi {\rm (ID)}$. The CD background contributes a roughly constant
fraction of the SD signal at the level of around 10\%.
The remaining DD and ND background sources are at or below the 1\% level.
 
The quality of the description of the backgrounds from the two largest sources is investigated
using control samples,
defined similarly to the main analysis selection, except that exactly two ALFA armlets
are required to contain a reconstructed proton, rather than one.
In `Control Region 1',
the remainder of the selection is as for the main analysis
(i.e.\ requiring activity in at least five MBTS sectors), which provides
a test of the overlay background treatment. In this case, the
two armlets containing protons are
in the back-to-back
azimuthal configuration
approximately 96\% of the time, indicating that elastic scattering is the dominant source
of ALFA background
signals. Figure~\ref{fig:controlPlotsTuned}(e) shows an example
($\Delta \eta$) control distribution in this sample.
The data are well described, with the proton overlay contribution heavily dominant and the CD contribution
being the next largest contribution.
In `Control Region 2',
the CD contribution is enhanced by requiring
activity in no fewer than two and no more than ten MBTS sectors.
The normalisation of this sample is well described, as are the shapes of all relevant distributions
except for that in $\xi {\rm (ID)}$. The $\xi$ dependence in the CD MC simulation is therefore reweighted to
better match the data in the control region
whilst preserving the normalisation,
yielding the description
shown in Figure~\ref{fig:controlPlotsTuned}(f).
The CD contribution is
the largest, although a substantial overlay background component remains.
The CD contribution can be
further enhanced by making even tighter requirements on small numbers of active MBTS sectors, at the expense
of accepting fewer events. The quality of the description remains at a level similar to that of Control Region 2.
 
\section{Unfolding}
The binning choices in the measurement are driven by the resolutions in each
variable, as quoted in Section~\ref{sec:fiducial}, such that the bin
purities\footnote{Bin purity is defined in the context of the simulation
as the fraction of all events reconstructed in a
measurement interval that are classified as SD and are also generated in that
interval at the generator level.}
are typically larger than $50 \%$.
After the background contributions are subtracted
and the trigger and ALFA efficiencies are accounted for,
the data are corrected for migrations between bins and
across the
fiducial boundaries of the measurement using
an iterative Bayesian unfolding
algorithm~\cite{DAGOSTINI1995487} based on the SD MC sample, reweighting the
input at each iteration.
The chosen number of iterations is a compromise between residual influence from the MC
generator-level prior (small numbers of iterations) and exaggeration of statistical effects (large numbers).
The optimal choice is determined by minimisation of the unfolding
systematic uncertainty (see Section~\ref{sec:systematics}) and varies between 1 and 4 for the
distributions studied.
The response matrices for the $|t|$ and $\Delta\eta$ variables are diagonal to a good approximation;
the response matrix for $\xi$ is also approximately diagonal after the correction for unreconstructed
particles described in section~\ref{sec:fiducial}.

 
\section{Uncertainties}
\label{sec:systematics}
The largest contribution to the systematic uncertainty
in many of the measurement bins arises from the overlay background subtraction.
This uncertainty is derived from the bin-by-bin fractional
difference between the data and the data-driven prediction
in Control Region 1,
propagated to the main selection.
It is less than
5\% in most bins, but grows to almost 20\% at the smallest and largest values of $|t|$.
 
The assumed ratios of the SD, DD and CD cross sections
enter the measurement through the background subtraction procedures. The ranges of
systematic variation are
chosen to match measurements by CDF \cite{Abe:1993wu,Affolder:2001vx,Acosta:2003xi},
which are compatible with the study of Control Region 2.
The assumed CD cross section is varied between
$1.12 \ {\rm mb}$ and $1.66 \ {\rm mb}$, which results in an uncertainty at the $5\%$ level.
The shape of the CD $\xi$ distribution is also altered in the MC simulation
to improve the description of the data as described
in Section~\ref{sec:controlPlots}. The associated systematic uncertainty is taken from the
difference between the unfolded results obtained when applying this reweighting and those obtained using the
original \pythia{} distribution. This difference results in uncertainties of up to 2\%.
Systematic variation of the DD cross section (between 29\% and 68\% of the SD cross section) leads to a
negligible uncertainty.
 
The systematic uncertainty arising from the unfolding is determined
via a `closure' test, in which the
reconstructed (detector level) \pythia{} A3 MC
distributions are first reweighted using high-order polynomials
to provide a close match to the background-subtracted detector-level data, and
are then unfolded using the same MC model with no reweighting applied. The
uncertainty is taken to be the fractional non-closure,
i.e.\ the deviation of the unfolded distributions from the generator-level distributions.
The resulting uncertainties
reach 5\% in the $\xi$ distribution and 2\% in the $\Delta \eta$ distribution.
Beyond this non-closure unfolding uncertainty, a further `model-dependence' uncertainty
arises from the simulation of the hadronisation of the system $X$.
This is evaluated by comparing the response predicted in
\pythia~with that from \herwig{}. It amounts to around $5\%$ in the $\xi$ measurement and is
negligible for $\Delta \eta$ and $t$.
 
The ALFA alignment and reconstruction uncertainties are
obtained using the methods described in Ref.~\cite{Aad:2014dca}.
The `horizontal'
alignment gives rise to the largest effect, causing an uncertainty of
typically 1\% in the $|t|$ distribution.
The luminosity uncertainty is 1.5\%, as determined from van der Meer
scans \cite{Aaboud:2016ijx,Aaboud:2016hhf}.
Other systematic uncertainties considered include those due to
the ID track reconstruction efficiencies, obtained following the methods described in Ref.~\cite{Aad:2016xww};
the trigger efficiency, obtained by varying the reference trigger;
and residual MBTS noise, obtained by varying the threshold. None of these produce
uncertainties in the measured cross sections beyond the 2\% level.
 
The final systematic uncertainties are obtained by adding the upward and downward shifts
from all sources separately in quadrature and symmetrising by taking the larger of the two
shifts. Typically they amount to between 5\% and 10\%, except at the extremes
of the measurement range in $t$. Statistical
uncertainties from the
number of events in the SD candidate
selection are negligible, but they do arise from the data-driven
overlay background subtraction; these are added in quadrature with the systematic uncertainties
to compute the total uncertainty.
 
 
\section{Results}
\label{sec:result}
\begin{figure}[h]
\centering
\includegraphics[width=0.7\linewidth]{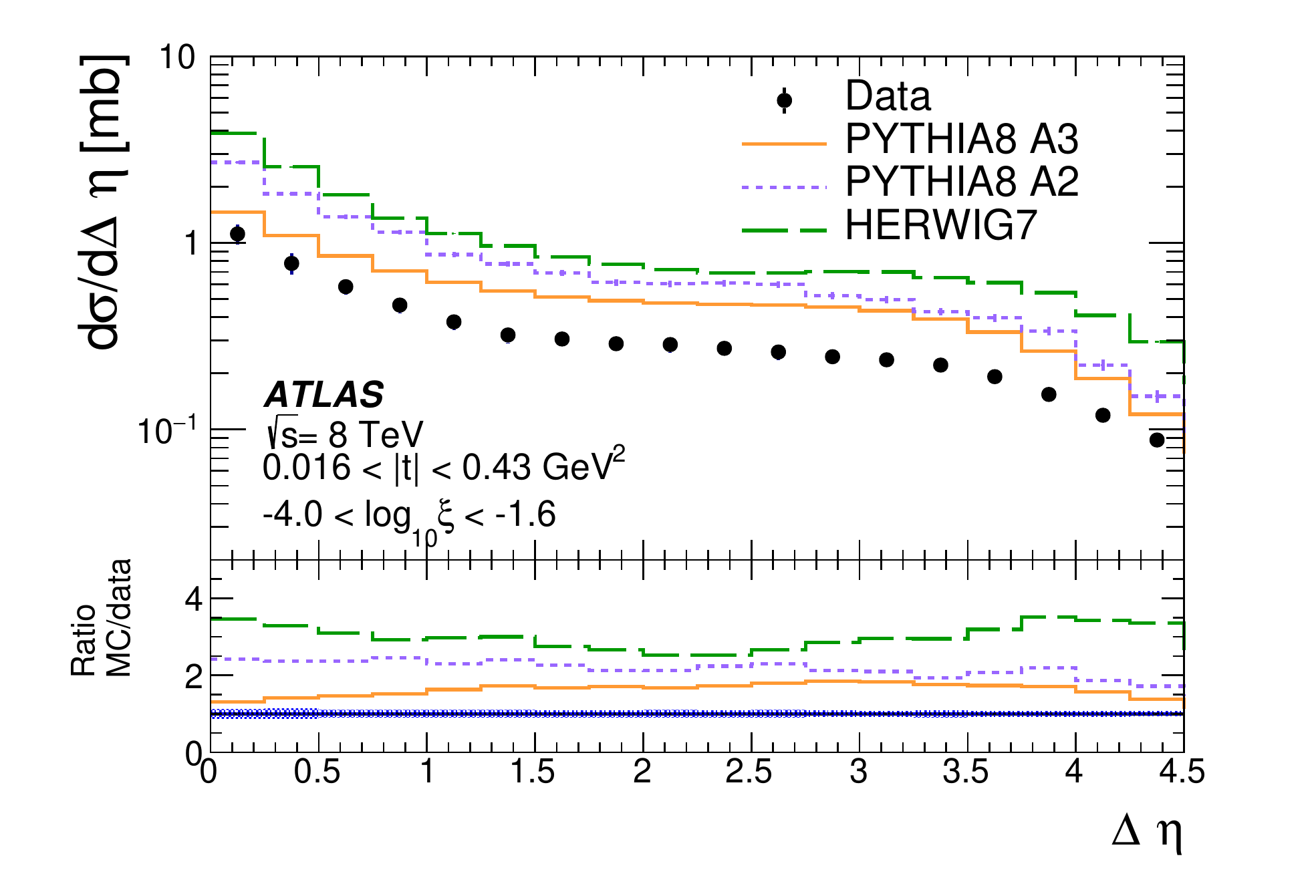}
\caption{Hadron-level differential SD cross section as a function of $\Delta\eta$,
comparing the measured data with \pythia~and \herwig{} predictions.
The error bars on the data points and the
band around unity in the MC/data ratio show the combination in quadrature
of fractional statistical and systematic uncertainties.}
\label{fig:crossSections}
\end{figure}
 
The background-subtracted, unfolded hadron-level SD cross sections are integrated over
the fiducial region $-4.0 < \log_{10}\xi < -1.6$ and
$0.016 < |t| < 0.43 \ \text{GeV}^2$ and correspond to cases where either of the two protons dissociates.
The differential cross section in $\Delta \eta$, defined in terms of primary charged particles
with $\pt > 200 \ {\rm MeV}$ as described in Section~\ref{sec:fiducial},
is shown in Figure~\ref{fig:crossSections}.
The error bars indicate the statistical and systematic uncertainties added in quadrature,
although the statistical contributions are negligible for most data points.
For gap sizes between about 1.5 and 3.5,
the differential cross section exhibits the plateau that
is characteristic of rapidity gap distributions in soft diffractive processes. There are
deviations from this behaviour at smaller and larger gap sizes due to the definition
of the observable in terms of a restricted rapidity region corresponding to the ID acceptance,
and to the fiducial range restriction, respectively.
The data are compared with the SD process simulations in the A2 and A3 tunes of \pythia, which exceed the measurement
by factors of 2.3 and 1.5, respectively. Both of these tunes are based on an integrated
SD cross section, defined according to the \pythia{} model, of $12.5 \ {\rm mb}$.
The difference between their predictions for the fiducial region
of the measurement arises from the different pomeron intercepts $\alpha(0)$ in their flux factors
(see Section~\ref{sec:MC}).
Both models give a reasonable description of the
shape of the $\Delta \eta$ distribution, the A2 tune being slightly better than A3.
The excess of the \pythia~prediction over data is compatible with previous ATLAS observations
from rapidity gap spectra \cite{Aad:2012pw} assuming the DD contribution to the \pythia~model
of the previous measurement is correct. The \herwig{} prediction is also broadly in line
with the shape of the
$\Delta \eta$ distribution, but exhibits an even larger excess in normalisation. This may
be partly due to the operational definition of the SD process that is adopted in the default SD model
normalisation, which is
derived from a rapidity gap measurement that also contains a DD admixture \cite{Gieseke:2016fpz}.

\begin{figure}[h] 
\centering
\includegraphics[width=0.7\linewidth]{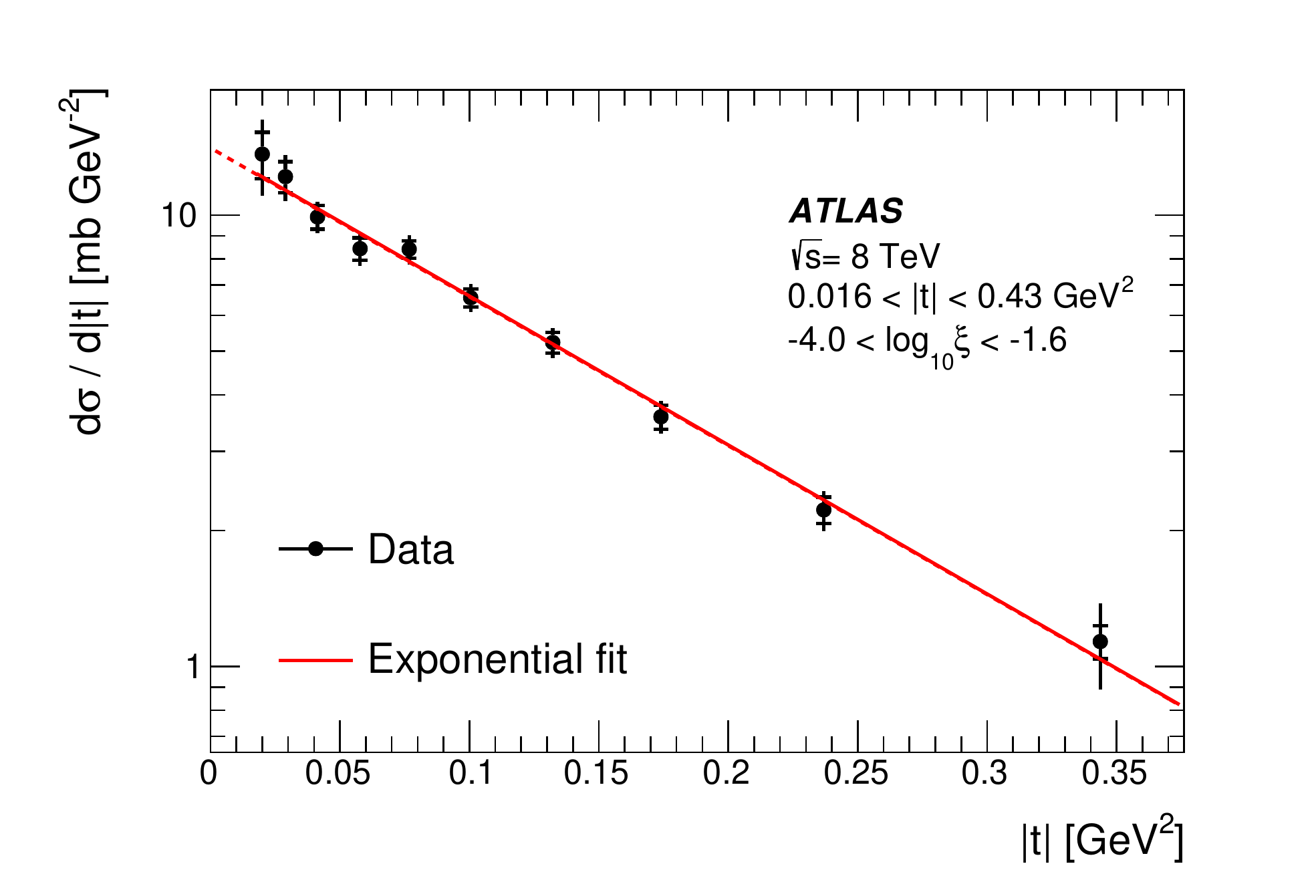}
\caption{The differential cross section as a function of $|t|$ with
inner error bars representing
statistical uncertainties and outer error bars displaying
the statistical and systematic uncertainties
added in quadrature. The result of the exponential fit described in the text is overlaid.}
\label{fig:fittedT}
\end{figure}
 
The cross section is shown differentially in $|t|$ in Figure~\ref{fig:fittedT}.
To avoid bias in the fit due to the fast-falling
nature of the distribution, the data points are plotted at the average values
of $|t|$ in each bin, as calculated from the corrected data.
The differential cross section is subjected to a fit of the form
${\text d} \sigma / {\text d} t \propto \text{e}^{Bt}$, which is overlaid on the figure.
The quality of the fit is acceptable ($\chi^2 = 8.3$ with eight degrees of freedom,
considering statistical uncertainties only). The result is
$B = 7.65 \pm 0.26 {\rm (stat.)} \pm 0.22 {\rm (syst.)} \ {\rm GeV^{-2}}$,
where the central value and statistical uncertainty are
obtained by fitting with statistical uncertainties only, and the
systematic uncertainty is obtained
by repeating the fit
separately for each systematic shift and adding the resulting deviations
from the central value in quadrature.
The measured slope parameter $B$
corresponds to a value averaged over the fiducial
$\xi$ range, with
$\langle \log_{10} \xi \rangle ~ = -2.88 \pm 0.14$,
where the central value is taken
from
the \pythia{} A3 tune and the uncertainty is defined by the difference from the \pythia{} A2 tune.
The largest contribution to the uncertainty in $B$ arises from the proton overlay
background subtraction, which has both a statistical and a systematic component.
The result is stable with respect to variations of the fitted $t$ range and is broadly
as expected from extrapolations of lower-energy measurements.
It is compatible with the predictions of $7.10 \ {\rm GeV^{-2}}$ from the Donnachie--Landshoff flux
and $7.82 \ {\rm GeV^{-2}}$ from Schuler--Sj\"{o}strand, contained in the \pythia{} A3 and A2
tunes, at the $1.6 \sigma$ and $0.5 \sigma$ levels, respectively.
 
\begin{figure}[h] 
\centering
\includegraphics[width=0.7\linewidth]{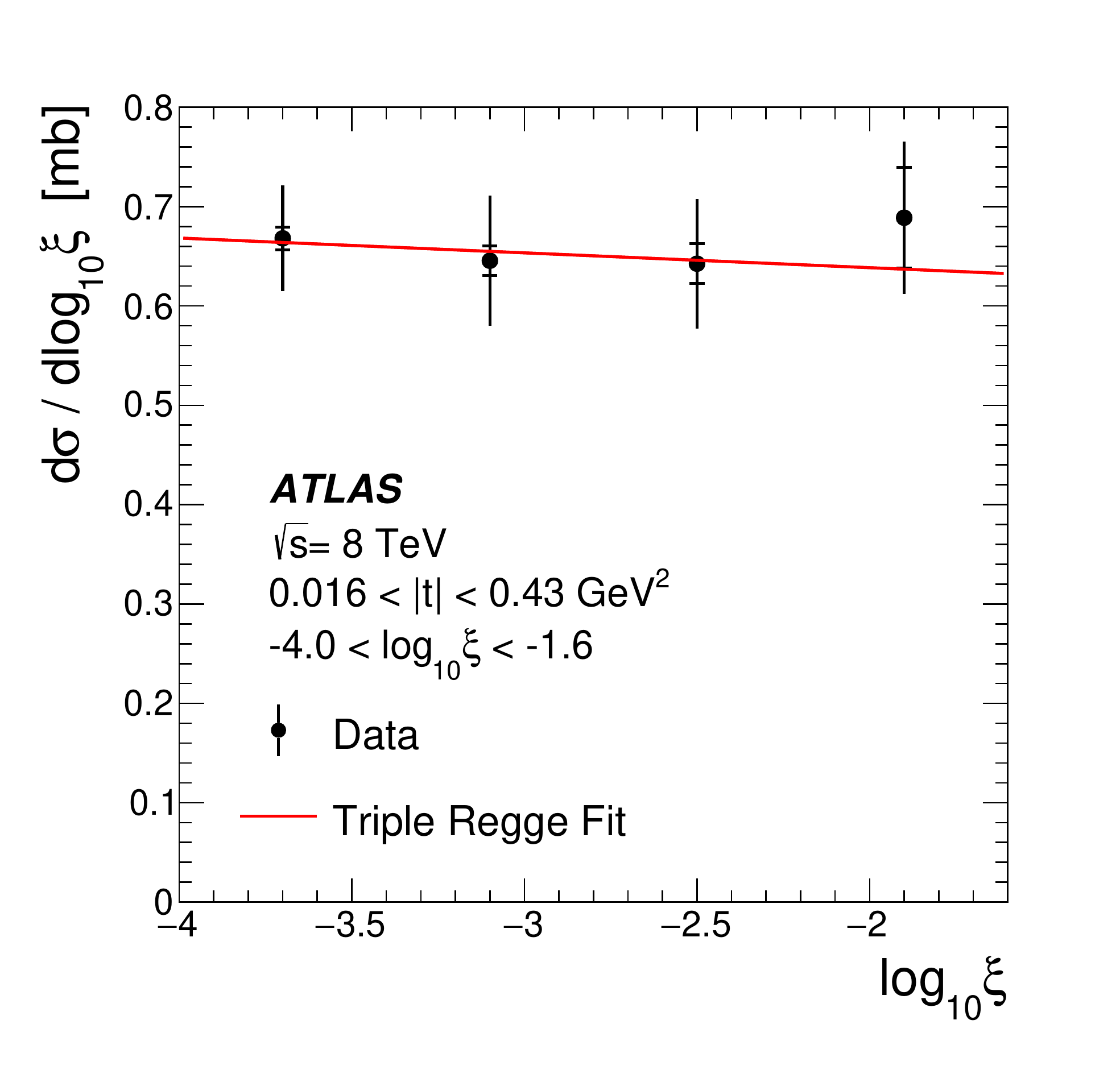}
\caption{The differential cross section as a function of $\log_{10} \xi$.
The inner error bars
represent only statistical uncertainties while the outer error bars display the
combination of statistical and systematic uncertainties in quadrature.
The data are compared with the results of the triple Regge fit described in the text.}
\label{fig:allXiFitted}
\end{figure}
 
~
 
In Figure~\ref{fig:allXiFitted},
the cross section is shown differentially in $\log_{10} \xi$, as obtained from the charged particles
reconstructed in the ID. Fully compatible results are obtained when reconstructing $\xi$ using ALFA,
despite the fast-deteriorating resolution at small $\xi$ values
and
completely different systematic effects. The data are compatible with being independent of this variable,
characteristic of the expected
behaviour of the cross section roughly as ${\text d} \sigma / {\text d} \xi \sim 1 / \xi$.
A more detailed interpretation of the $\xi$
dependence
is obtained
through a fit to the data in the framework of Regge phenomenology.
At asymptotically large fixed $s$,
and with $s \gg M_X^2 \gg |t|$,
the double-differential cross section in $\xi$ and $t$ is expected to follow the `triple Regge'
form \cite{Adloff:1997mi,Kaidalov:1979jz,1970Mueller,ALBERI19811,GOULIANOS1983169,0038-5670-31-2-R02},
\begin{eqnarray*}
\frac{{\text d}^2 \sigma}{{\text d} \xi {\text d} t} \propto
\left( \frac{1}{\xi} \right)^{2 \alpha(t) - 1} \ (M_{X}^{2})^{\alpha(0)-1} \ \text{e}^{B_0 t} \ .
\end{eqnarray*}
Here, the first factor on the right hand side represents the pomeron flux factor,
the second factor corresponds to the total pomeron--proton cross
section\footnote{This $M_X$-dependent term, deriving from Mueller's
generalisation of the optical theorem \cite{1970Mueller}, is
commonly treated differently, particularly in models that attempt to make the link to partonic behaviour
and QCD. For example in \pythia, it is taken to be constant. Neglecting  this contribution
leads to a decrease in the extracted $\alpha(0)$ in the current analysis by 0.03.}
and the exponential $t$ dependence is empirically motivated, $B_0$ characterising the spatial size
of the scattering protons.
Integrating over the fiducial $t$ range of the measurement between
$t_{\rm low} = -0.43 \ {\rm GeV^2}$ and $t_{\rm high} =  -0.016 \ {\rm GeV^2}$
yields a prediction for the single-differential cross section
\begin{eqnarray}
\frac{{\text d} \sigma}{{\text d} \xi} \propto
\left( \frac{1}{\xi} \right)^{\alpha(0)} \ \frac{\text{e}^{B t_{\rm high}} - \text{e}^{B t_{\rm low}}}{B} \ ,
\label{trb:equation}
\end{eqnarray}
where the $t$ dependence of the pomeron trajectory is absorbed into
$B = B_0 - 2 \alpha^\prime \ln \xi$.
In this type of model, the $\xi$ dependence therefore measures the value of the pomeron intercept.
A fit of the form of Eq.~(\ref{trb:equation}) is
applied to the measured $\xi$ distribution
with $\alpha(0)$ and the overall normalisation
as free parameters. The Donnachie--Landshoff value for the slope of the pomeron trajectory
$\alpha^\prime = 0.25 \ {\rm GeV^{-2}}$ is taken for the central value, with
$\alpha^\prime = 0$ used to determine the associated uncertainty.
This fit, displayed in Figure~\ref{fig:allXiFitted}, yields a value of
$\alpha(0) = 1.07 \pm 0.02 \ {\rm (stat.)} \pm 0.06 \ {\rm (syst.)} \pm 0.06 \ {\rm (\alpha^\prime)}$.
The largest systematic uncertainties apart from the $\alpha^\prime$ assumption
arise from the unfolding, the hadronisation uncertainty and the overlay background subtraction.
This result is compatible with predictions using
soft pomeron phenomenology and assuming a universality between total, elastic, and diffractive
cross sections.
It can be compared with the predictions of $1.14$ and $1.00$ from the \pythia~A3 and A2 tunes,
respectively,
when applying the triple Regge formalism in place of the default \pythia~model to which the A3 input value of $1.07$
is applicable.
It is not possible to compare the extracted $\alpha(0)$ and $B$ parameters
with predictions from \herwig{}, since the $\xi$ dependence of the $B$ slope has a complex behaviour
in that model.

The cross section integrated over the full fiducial range of the analysis,
$-4.0 < \log_{10}\xi \leq -1.6$ and $0.016 < |t| \leq 0.43$ GeV$^{2}$,
is $1.59 \pm 0.03 \ {\rm (stat.)} \pm \ 0.13 \ {\rm (syst.)} \ {\rm mb}$, with the largest contribution
to the uncertainty arising from the proton overlay subtraction.
Extrapolating to the full $t$ range assuming the measured slope parameter $B$
leads to a cross section of $1.88 \pm 0.15 \ {\rm mb}$ integrated over $-4.0 < \log_{10}\xi \leq -1.6$,
with statistical and systematic uncertainties combined.
The cross sections before and after this extrapolation are compared with predictions from the MC models
in Table \ref{table:allXses}.
 
\begin{table}[h!]
\caption{The SD cross section within the fiducial region
($-4.0 < \log_{10}\xi \leq -1.6$ and $0.016 < |t| \leq 0.43$ GeV$^{2}$) and
extrapolated across all $t$ using the measured slope parameter $B$.
The systematic and statistical uncertainties are combined for data.
The MC statistical uncertainties are negligible.}
\centering
\begin{tabular}{c|c|c}
\toprule
Distribution & $\sigma_{\rm SD}^{\text{fiducial}(\xi,t)}$ [mb] & $\sigma_{\rm SD}^{t\text{-extrap}}$ [mb] \\ \midrule
Data                                 & $1.59\pm0.13$ & $1.88\pm0.15$ \\
\pythia{} A2 (Schuler--Sj\"{o}strand) & $3.69$ & $4.35$ \\
\pythia{} A3 (Donnachie--Landshoff)   & $2.52$ & $2.98$ \\
\herwig{}                             & $4.96$ & $6.11$ \\
\bottomrule
\end{tabular}
\label{table:allXses}
\end{table}

The behaviour of the SD cross section at $\xi$ values beyond
the measured region is not yet well constrained by LHC data, and phenomenological models
predict additional terms at both extremes (e.g.\ involving sub-leading exchanges in the Regge case).
It is therefore not possible to make a reliable assessment of the uncertainties inherent in extrapolating to a
full SD cross section. However, an estimate
of the corresponding parameter in the \pythia{} model can be obtained,
assuming that the excess of MC over data in the fiducial
region studied persists throughout the full kinematic range.
Since the measurement of $\alpha(0)$ lies midway between the predictions
of the A3 and A2 tunes of \pythia{}, the estimate is obtained
by scaling the measured fiducial cross section
by the average of the extrapolation factors predicted by the two tunes.
The total SD cross-section parameter in the \pythia{} model then decreases from
\SI{12.5}{mb} to \SI{6.6}{mb}.

\section{Summary}
A detailed study is performed of the
dynamics of the inclusive single-diffractive dissociation process $pp \rightarrow Xp$
at $\sqrt{s} = 8 \ {\rm TeV}$ using the
ATLAS detector at the LHC.
Unlike in previous related analyses, the final-state protons are reconstructed directly,
using the ALFA forward spectrometer.
Differential cross sections are measured as a function of the fractional proton energy
loss $\xi$, the
squared four-momentum transfer $t$, and the size $\Delta \eta$ of the pseudorapidity
interval on the same side of the interaction point as the intact proton
extending from $\eta = \pm 2.5$ to the closest charged particle with smaller $|\eta|$ and  $\pt > 200 \ {\rm MeV}$.
The fiducial range of the measurement is $-4.0 < \log_{10} \xi < -1.6$
and $0.016 < |t| < 0.43 \ {\rm GeV^2}$.
For gap sizes between approximately 1.5 and 3.5,
the cross section differential in $\Delta \eta$ exhibits the plateau that
is characteristic of rapidity gap distributions in soft diffractive processes. There are
deviations from the plateau at larger and smaller gap sizes due to the definition
of the observable and the acceptance.
The cross section differential in $t$ is well described by an exponential behaviour,
${\rm d} \sigma / {\rm d} t \propto \text{e}^{Bt}$ with the slope parameter measured to be
$B = 7.65 \pm 0.34 \ {\rm GeV^{-2}}$, consistent with expectations and
with extrapolations
from lower-energy measurements.
The variable $\xi$ is reconstructed using two complementary methods, based on either the
scattered proton in ALFA or the tracks in the ID. The ID-track-based measurement is adopted and the standard
triple pomeron approach of Regge phenomenology
is used to describe the data in terms of
a pomeron trajectory with intercept $\alpha(0) = 1.07 \pm 0.09$, in
good agreement with previous values
from ATLAS and elsewhere.
The measured cross section integrated over the fiducial region
amounts to $1.59 \pm 0.13 \ {\rm mb}$. This is
substantially smaller than is predicted in the tunes of \pythia{}
and, particularly, \herwig{} that were used in the analysis.


\section*{Acknowledgements}
 

We thank CERN for the very successful operation of the LHC, as well as the
support staff from our institutions without whom ATLAS could not be
operated efficiently.
 
We acknowledge the support of ANPCyT, Argentina; YerPhI, Armenia; ARC, Australia; BMWFW and FWF, Austria; ANAS, Azerbaijan; SSTC, Belarus; CNPq and FAPESP, Brazil; NSERC, NRC and CFI, Canada; CERN; ANID, Chile; CAS, MOST and NSFC, China; COLCIENCIAS, Colombia; MSMT CR, MPO CR and VSC CR, Czech Republic; DNRF and DNSRC, Denmark; IN2P3-CNRS and CEA-DRF/IRFU, France; SRNSFG, Georgia; BMBF, HGF and MPG, Germany; GSRT, Greece; RGC and Hong Kong SAR, China; ISF and Benoziyo Center, Israel; INFN, Italy; MEXT and JSPS, Japan; CNRST, Morocco; NWO, Netherlands; RCN, Norway; MNiSW and NCN, Poland; FCT, Portugal; MNE/IFA, Romania; JINR; MES of Russia and NRC KI, Russian Federation; MESTD, Serbia; MSSR, Slovakia; ARRS and MIZ\v{S}, Slovenia; DST/NRF, South Africa; MICINN, Spain; SRC and Wallenberg Foundation, Sweden; SERI, SNSF and Cantons of Bern and Geneva, Switzerland; MOST, Taiwan; TAEK, Turkey; STFC, United Kingdom; DOE and NSF, United States of America. In addition, individual groups and members have received support from BCKDF, CANARIE, Compute Canada, CRC and IVADO, Canada; Beijing Municipal Science \& Technology Commission, China; COST, ERC, ERDF, Horizon 2020 and Marie Sk{\l}odowska-Curie Actions, European Union; Investissements d'Avenir Labex, Investissements d'Avenir Idex and ANR, France; DFG and AvH Foundation, Germany; Herakleitos, Thales and Aristeia programmes co-financed by EU-ESF and the Greek NSRF, Greece; BSF-NSF and GIF, Israel; La Caixa Banking Foundation, CERCA Programme Generalitat de Catalunya and PROMETEO and GenT Programmes Generalitat Valenciana, Spain; G\"{o}ran Gustafssons Stiftelse, Sweden; The Royal Society and Leverhulme Trust, United Kingdom.
 
The crucial computing support from all WLCG partners is acknowledged gratefully, in particular from CERN, the ATLAS Tier-1 facilities at TRIUMF (Canada), NDGF (Denmark, Norway, Sweden), CC-IN2P3 (France), KIT/GridKA (Germany), INFN-CNAF (Italy), NL-T1 (Netherlands), PIC (Spain), ASGC (Taiwan), RAL (UK) and BNL (USA), the Tier-2 facilities worldwide and large non-WLCG resource providers. Major contributors of computing resources are listed in Ref.~\cite{ATL-SOFT-PUB-2020-001}.
 

\printbibliography

\clearpage
 
\begin{flushleft}
{\Large The ATLAS Collaboration}

\bigskip

G.~Aad$^\textrm{\scriptsize 101}$,    
B.~Abbott$^\textrm{\scriptsize 128}$,    
D.C.~Abbott$^\textrm{\scriptsize 102}$,    
O.~Abdinov$^\textrm{\scriptsize 13,*}$,    
A.~Abed~Abud$^\textrm{\scriptsize 70a,70b}$,    
K.~Abeling$^\textrm{\scriptsize 53}$,    
D.K.~Abhayasinghe$^\textrm{\scriptsize 93}$,    
S.H.~Abidi$^\textrm{\scriptsize 167}$,    
O.S.~AbouZeid$^\textrm{\scriptsize 40}$,    
N.L.~Abraham$^\textrm{\scriptsize 156}$
% ...
\bigskip
\\

% ...
$^{13}$Institute of Physics, Azerbaijan Academy of Sciences, Baku; Azerbaijan.\\
% ...
$^{40}$Niels Bohr Institute, University of Copenhagen, Copenhagen; Denmark.\\
% ...
$^{53}$II. Physikalisches Institut, Georg-August-Universit\"{a}t G\"ottingen, G\"ottingen; Germany.\\
% ...
$^{70}$$^{(a)}$INFN Sezione di Pavia;$^{(b)}$Dipartimento di Fisica, Universit\`a di Pavia, Pavia; Italy.\\
% ...
$^{93}$Department of Physics, Royal Holloway University of London, Egham; United Kingdom.\\
% ...
$^{101}$CPPM, Aix-Marseille Universit\'e, CNRS/IN2P3, Marseille; France.\\
$^{102}$Department of Physics, University of Massachusetts, Amherst MA; United States of America.\\
% ...
$^{128}$Homer L. Dodge Department of Physics and Astronomy, University of Oklahoma, Norman OK; United States of America.\\
% ...
$^{156}$Department of Physics and Astronomy, University of Sussex, Brighton; United Kingdom.\\
% ...
$^{167}$Department of Physics, University of Toronto, Toronto ON; Canada.\\
% ...
$^{a}$ Also at Borough of Manhattan Community College, City University of New York, New York NY; United States of America.\\
$^{b}$ Also at CERN, Geneva; Switzerland.\\
% ...
$^{*}$ Deceased

\end{flushleft}


\end{document}